\documentclass[12pt]{article}

\usepackage[pdftex]{graphicx}
\usepackage{amssymb,amsfonts,amsmath}
\usepackage{cite}
\usepackage[auth-sc]{authblk}
\usepackage[top=1in, bottom=1in, left=1in, right=1in]{geometry}
\usepackage{hyperref}

\usepackage[doublespacing]{setspace}
\usepackage{lineno}

\newlength{\HalfPage}
\begin{document}

\title{Exploring the evolution of a trade-off between vigilance and foraging in group-living organisms}

\author[1,5,*]{\small Randal S.~Olson}
\author[2,5,*]{Patrick B.~Haley}
\author[3,5]{Fred C.~Dyer}
\author[4,5]{Christoph Adami}

\affil[1]{Computer Science \& Engineering Dept., Michigan State University, East Lansing, MI 48824}
\affil[2]{Computer Science Dept., The University of Texas at Austin, Austin, TX 78712}
\affil[3]{Zoology Dept., Michigan State University, East Lansing, MI 48824}
\affil[4]{Microbiology \& Molecular Genetics Dept., Michigan State University, East Lansing, MI 48824}
\affil[5]{BEACON Center for the Study of Evolution in Action, East Lansing, MI 48824}
\affil[*]{These authors contributed equally to this work}
\affil[ ]{E-mail: olsonran@msu.edu, patrick.haley@utexas.edu, fcdyer@msu.edu, adami@msu.edu}

\date{}

\maketitle

{\parindent0pt Keywords: {\it group foraging}, {\it many eyes hypothesis}, {\it anti-predator vigilance}, {\it genetic relatedness}, {\it reproductive strategy}, {\it tragedy of the commons}}

\hspace{0.1in}

{\parindent0pt Article submitted to the Journal of the Royal Society Interface, August 2014.}


\newpage

\section*{Abstract}

Despite the fact that grouping behavior has been actively studied for over a century, the relative importance of the numerous proposed fitness benefits of grouping remain unclear. We use a digital model of evolving prey under simulated predation to directly explore the evolution of gregarious foraging behavior according to one such benefit, the ``many eyes'' hypothesis. According to this hypothesis, collective vigilance allows prey in large groups to detect predators more efficiently by making alarm signals or behavioral cues to each other, thereby allowing individuals within the group to spend more time foraging. Here, we find that collective vigilance is sufficient to select for gregarious foraging behavior as long there is not a direct cost for grouping (e.g., competition for limited food resources), even when controlling for confounding factors such as the dilution effect. Further, we explore the role of the genetic relatedness and reproductive strategy of the prey, and find that highly related groups of prey with a semelparous reproductive strategy are the most likely to evolve gregarious foraging behavior mediated by the benefit of vigilance. These findings, combined with earlier studies with evolving digital organisms, further sharpen our understanding of the factors favoring grouping behavior.

\newpage

\section{Introduction}

Many prey choose to live, forage, and reproduce in groups---this is one of the most readily-observed phenomena in biology. A common adaptive explanation for grouping behavior is that it aids in anti-predatory defense. For instance, Starlings ({\it Sturnus vulgaris}) are well known to forage in flocks in the presence of predators~\cite{Powell1974}. Shoaling fish, e.g. the eastern mosquitofish ({\it Gambusia holbrooki}), have been documented to identify predators more accurately in larger groups~\cite{Ward2011}. Ostriches ({\it Struthio camelus}) have been reported to experience anti-predatory benefits when foraging in groups~\cite{Bertram1980}. Even when there is a correlation between grouping behavior and protection from predators, however, it is difficult to pin down what benefits actually select for the evolution of grouping behavior.

Several such fitness benefits have been proposed. For example, grouping can improve group vigilance~\cite{Pulliam1973,Treisman1975,Kenward1978,Treherne1981}, reduce the chance of being encountered by predators~\cite{Treisman1975,Inman1987}, dilute an individual's risk of being attacked~\cite{Hamilton1971,Foster1981,Treherne1982,Ioannou2012,Olson2013SelfishHerd,Olson2014SelfishHerd}, enable an active defense against predators~\cite{Bertram1978}, or reduce predator attack efficiency by confusing the predator~\cite{Krakauer1995,Kunz2006PredatorConfusion,Jeschke2007,Ioannou2008,Olson2013PredatorConfusion}. Other possible benefits not involving predation include improved mating success~\cite{Yuval1993}, increased foraging efficiency~\cite{Pulliam1984}, and the ability for the group to solve problems that would be impossible to solve individually~\cite{Couzin2009}, for example through the division of labor~\cite{Goldsby2011}.

With all of these interdependent factors potentially affecting the evolution of grouping, it is difficult to study the independent effects of each benefit in biological systems, let alone explore how they unfold over evolutionary time scales. However, recent research has shown that it is possible to explore the potential independent effects of each benefit by modeling them with digital models of evolution~\cite{Hamblin2013}. In previous work, we created several models to explore the predator confusion~\cite{Olson2013PredatorConfusion} and selfish herd~\cite{Olson2013SelfishHerd,Olson2014SelfishHerd} hypotheses to find when these benefits do (and do not) independently select for grouping behavior. One advantage of these models is that once the independent effects of the various grouping benefits are understood, we can then combine the benefits into a single model to study their relative importance and separate the adaptive benefits (that select for the evolution of grouping) from the incidental side effects of grouping.

Here, we focus on anti-predator vigilance (i.e., the many eyes hypothesis) as a possible selective mechanism for the evolution of gregarious foraging behavior, and control for the influence of the other benefits described above. First proposed using a mathematical model~\cite{Pulliam1973} and explored experimentally a year later~\cite{Powell1974}, the many eyes hypothesis makes two key predictions, both of which arise from the assumption that vigilance is costly because it imposes a trade-off with foraging efficiency: (a) individual prey vigilance will decline as a group size increases, and (b) because prey can more equitably divide the task of watching for predators in large groups, they will experience a fitness benefit from foraging more. Therefore, there will be a selective advantage for prey that forage in groups up to a certain group size. In the 40 years since its inception, these predictions have been examined in numerous species across hundreds of independent studies~\cite{Caraco1979,Ward2011,Bertram1980,Roberts1996,Beauchamp2008}. Furthermore, several game theoretical models have been applied to refine the predictions of when collective vigilance in foraging groups should evolve~\cite{Pulliam1982}, and subsequently matched to experimental data~\cite{McNamara1992}.

These previous studies focus on the potential fitness consequences of vigilance in groups of animals, but they do not address the circumstances under which vigilance, and the advantages of being in a group with many watchful eyes, provides a sufficient selection pressure to favor group living, independent of other pressures. When considering the evolution of grouping behavior, it is vital to take into account both the benefits {\em and} costs imposed by the behavior~\cite{Clark1986}. To satisfy this requirement, a handful of researchers have recently turned to digital models to study the evolution of animal behavior~\cite{Beauchamp2007,Ruxton2008,Katsnelson2011}. These researchers use a digital model of evolution to evolve the behavior of a population of locally-interacting animats, enabling them to explore the evolution of behavior in complex environments that are beyond the means of mathematical models~\cite{Adami2012,Hamblin2013}. Additionally, these evolutionary model systems allow researchers to explicitly control for complicating factors, such as the dilution effect~\cite{Roberts1996} and food density~\cite{Elgar1989}, that are commonly confounded with collective vigilance as factors benefiting group-living organisms.

In this study, we extend these digital evolution models to explore the conditions under which collective vigilance favors the evolution of gregarious foraging behavior. We assume that vigilance has benefits (e.g., communicating the presence of a predator via alarm signals) but also costs (e.g., reduced foraging rates by watching for the predator). Under the many eyes hypothesis, grouping is beneficial because it reduces the cost of vigilance by sharing the cost of vigilance among the group, but it may have additional costs that must be considered, e.g., increased predation rates on larger groups~\cite{Ale2007}. Furthermore, this benefit would be diluted if some individuals can freeload on the vigilance of others (as in heterogeneous groups), but magnified if the group members are highly related. The benefits and costs would also be affected by the life history of the prey, in particular whether their reproduction is iteroparous (i.e., repeated) or semelparous (i.e., all at once): Vigilance may be more beneficial in semelparous prey because a predation event can completely prevent them from reproducing, whereas iteroparous prey are more likely to have reproduced at least once prior to experiencing a predation event. To explore these issues, we manipulate the genetic relatedness and reproductive strategy of groups of prey that are under predation and observe the resulting behavior after thousands of generations of digital evolution have taken place. A preliminary investigation of this work was published in the ALIFE 14 conference~\cite{Haley2014}, and has been significantly extended in this paper.

\section{Methods}

\begin{figure*}
\begin{center}
\includegraphics[width=0.9\textwidth]{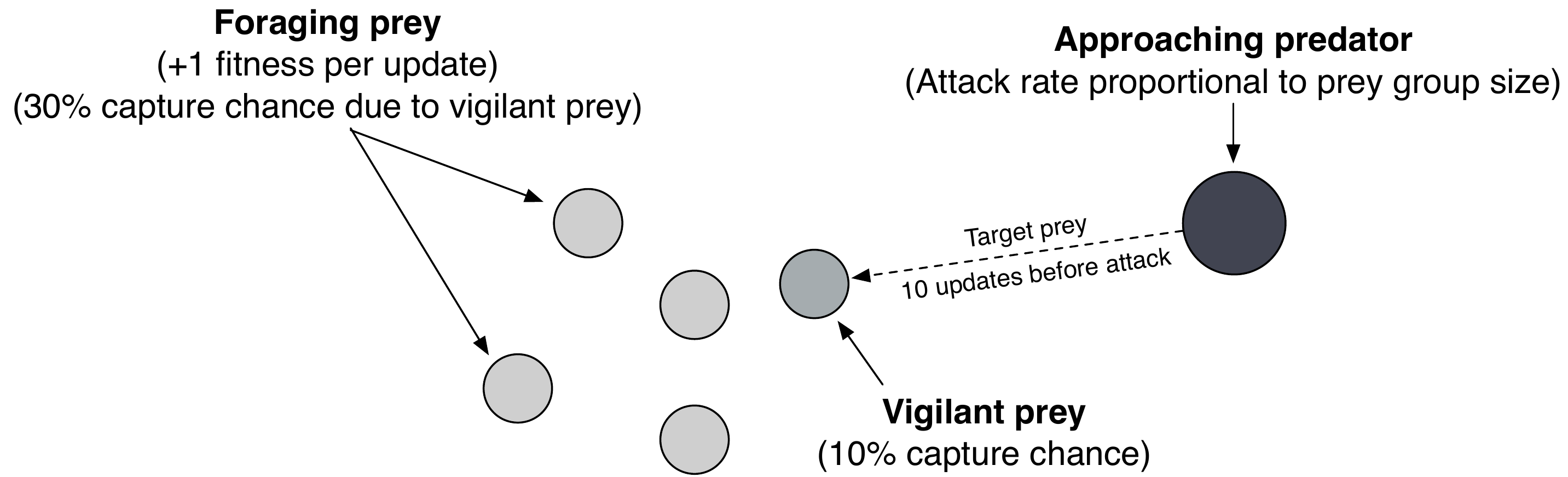}
\caption{{\bf Depiction of the disembodied simulation.} Prey seek to forage as much as possible while avoiding being captured by the predator. If none of the prey in the group are vigilant, the target prey is captured 100\% of the time.}
\label{fig:disembodied-sim}
\end{center}
\end{figure*}

Figure~\ref{fig:disembodied-sim} depicts our model of predator-prey interactions in a disembodied model, wherein prey must balance the trade-offs between foraging and vigilance~\cite{Ruxton2008}. In an embodied model~\cite{PfeifferBongard2006}, every animat is situated in the world, perceives the world via its sensors, and can act on the world via behavioral or other responses. While embodied models offer more detail, they are also sensitive to implementation-specific details of the sensors and actuators, which can skew results. We therefore focus on a disembodied model\footnote{Model code: https://github.com/phaley/eos/tree/non-embodied} for the remainder of this study.

In this model, prey fitness is directly related to the amount of time it spends foraging, where a single round of foraging increases prey fitness by 1.0. However, prey vigilance determines whether a predator's attack on the prey is successful. These two options---foraging and vigilance---are assumed to be mutually exclusive. Thus, prey must evolve to maximize their food intake while remaining vigilant enough to survive the entire simulation, which is akin to the maximum possible life span of the prey.

\subsection*{Model of predators and prey}

We designed this model to capture certain features of natural predators and to control for potentially complicating factors. First, to ensure that predator attacks are not trivially predictable we simulate predators that attack at intervals that are normally distributed around a specific attack rate. Thus, predator attacks are randomly distributed throughout the 2,000-time-step duration of the simulation. To model the fact that larger groups of prey often attract more attacks from predators---a realistic cost of group living known as the {\em attraction effect}~\cite{Ale2007}---we scale this attack rate with the group size, such that the group experiences 5 predator attacks for every prey initially in the group over the course of the simulation. This scaling factor also allows us to control for the {\em dilution effect}, which has been suggested to allow prey to survive with lower vigilance levels in larger groups only because they are less likely to be the target of a predator's attack~\cite{Roberts1996,Beauchamp2003,Fairbanks2007}.

Each time a predator appears, we randomly select a target prey from the surviving prey of previous attacks. This is followed by a 10 time step delay between the appearance of the predator in the simulation and the actual attack, representing the time it takes for a predator to close the distance to the prey. It is during this time that prey vigilance becomes important. If the target prey is vigilant at any time during this interval, then it spots the predator and the attack only has a 10\% chance of success. If the target prey is not vigilant but one or more other prey in the group are vigilant, then the other prey communicate the presence of the predator via an alarm signal or other behavioral indicator and the predator will capture the target prey 30\% of the time. These probabilities are chosen based on analytical models of group vigilance~\cite{Ruxton2008} such that group vigilance is not as effective as individual vigilance, and models the imperfect communication between members of the group~\cite{Lima1995b}. Finally, if no members of the group are vigilant while the predator is closing the distance to its target, then the entire group is unaware of the predator and the attack will succeed 100\% of the time. In all cases of a successful attack, the target prey is removed from the simulation and can no longer forage to increase its fitness.

Each individual prey makes the decision to forage or be vigilant every simulation time step. This decision-making process is modeled with a {\em Markov Network} (MN), which is an ``artificial brain'' that can stochastically make decisions based on sensory input, memory, and previous actions~\cite{Olson2013PredatorConfusion,Edlund2011,Marstaller2013}. Every prey MN is encoded by a list of numbers known as its genotype, such that changes to the genotype can result in changes in the function of the MN. Because we do not provide any sensory input to the prey in this simulation, we are effectively modeling the probability of a prey taking an action every simulation time step. More information on MNs---including details on their genetic encoding, mutational operators, and functionality---is available in~\cite{Olson2014SelfishHerd}.

\subsection*{Evolutionary process}

We repeat the evaluation procedure described above until all 100 individuals in the Genetic Algorithm (GA) population have been assigned a fitness (see, e.g.,~\cite{Eiben2003} for a full description of GAs). Once all individuals have been assigned a fitness, we use fitness-proportional selection according to a Moran process~\cite{Moran1962} to produce the next generation's population of prey. Fitness-proportional selection ensures that prey with higher fitness values generally produce more offspring. The selected prey reproduce asexually, with a small probability of mutations (0.5\% per site) affecting their offspring's genotype. We repeat this evaluation-selection-reproduction process for 2,500 generations to ensure that the GA has reached an evolutionarily stable strategy~\cite{Hamblin2007} and replicate the experiments 100 times for each treatment---each with a distinct random number generator seed---to verify that we are capturing evolutionary trends rather than outlier scenarios.

\subsection*{Genetic relatedness}

Since the many eyes hypothesis predicts an inverse relationship between individual vigilance and group size~\cite{Pulliam1973,Powell1974}, we study prey populations across a range of group sizes (5, 10, 25, and 50). In our first experiment, we observe the equilibrium vigilance levels when prey are forced to group. In the next experiment, we relax this assumption and allow the prey to choose to group (or not) every time step. In the latter case, we report the group size as the maximum initial group size. To provide a baseline for the optional grouping experiment, we compare its equilibrium vigilance levels to that of experiments where prey are forced to group and experiments where prey are forced to forage individually.

For all of the above experiments, we study the effect of genetic relatedness on grouping behavior. In these treatments, groups can be formed in two different ways. In homogeneous groups, each individual in the current generation of the GA's population is evaluated separately. During an individual's fitness evaluation, we fill the group with exact copies of the individual, and the fitness for that individual is the average fitness of all of its copies at the end of the simulation. Because vigilance indirectly benefits the vigilant individual in homogeneous groups by aiding its kin, we expect that vigilance will be highly beneficial in this treatment. In heterogeneous groups, we use a subset of the GA's population (which contains many prey with different genetics) to study how the prey fare in direct competition (or cooperation) with each other. Because the vigilance of one prey can potentially aid a rival prey in heterogeneous groups, we expect to observe lower levels of vigilance in this treatment.

\subsection*{Reproductive strategy}

The benefits of making the right decision in this simulated environment are straightforward: The prey must maximize food intake by surviving the longest while minimizing the time spent being vigilant. But the cost of making the wrong decision can also depend on the life history of the prey. For example, two different reproductive strategies---semelparity and iteroparity---should incur different costs. Semelparous organisms are characterized by a single reproductive event prior to death. We assume that this reproductive event occurs at the end of the simulation, so if a semelparous prey is consumed by the predator before the end of the simulation, all of its gathered food counts for nothing: it will leave no offspring. In contrast, iteroparous organisms continually reproduce throughout their lifetime. Therefore, when a predator consumes an iteroparous prey, the prey can no longer increase its fitness, but any food it gathered prior to its death counts toward its fitness for the simulation. We hypothesize that the increased risk of genetic death introduced by the semelparous treatment will provide an evolutionary incentive for prey to invest in vigilance, whereas prey in the iteroparous treatment will be more likely to engage in risky, non-cooperative behavior because their demise does not necessarily doom their genetic lineage~\cite{Hintze2014RiskAversion}.

\subsection*{Explicit cost of grouping}

The model described so far includes a cost of vigilance (insofar as prey cannot forage at the same time that they are vigilant), but there is no explicit cost to choosing to group aside from the possibility of aiding a competing individual. In a final treatment, we implement such a grouping penalty in order to model the realistic constraints of limited resources and the resulting scramble competition for food~\cite{Elgar1989,McNamara1992,Beauchamp2003,Sansom2008}. This grouping penalty is only assessed on prey who choose to forage in the group, and decreases the amount of food they receive in that simulation time step proportional to the number of prey in the group. The group foraging penalty is imposed according to the equation:
\begin{equation}
{\rm Food} = \frac{1.0}{M * G}
\end{equation}
where $G$ is the number of prey in the group and $M$ is the penalty multiplier that allows us to experimentally control the severity of the penalty. Given this penalty, prey foraging in larger groups receive less food every time they forage, but potentially enjoy the benefits of group vigilance.

\section{Results}

We evolve the vigilance behavior of prey by subjecting them to predation under a variety of treatments that vary reproductive strategy and group composition. Vigilance is measured as the percent chance that a prey will be vigilant at a given moment in time, averaged across all of the prey in the population. These treatments are repeated across a wide range of group sizes, allowing us to study not only whether the selection for vigilance can be generalized to groups of varying sizes, but also whether we can observe the inverse relationship between group size and vigilance predicted by the many eyes hypothesis.

\begin{figure*}
\begin{center}
\includegraphics[width=0.95\textwidth]{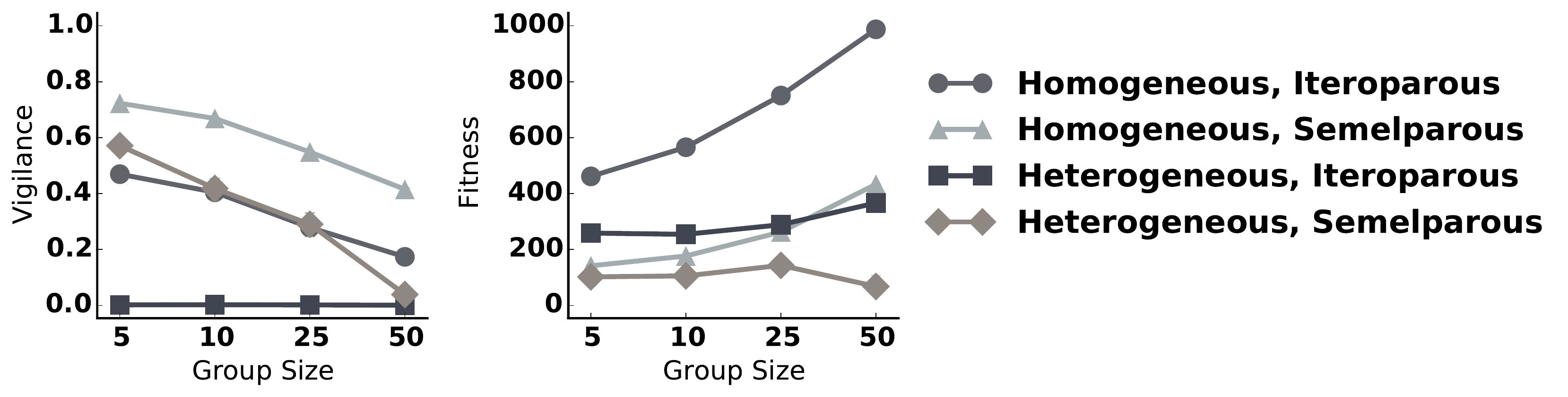}
\caption{{\bf Treatment comparison when prey are forced to forage in groups.} Both group homogeneity and a semelparous reproductive strategy select for high levels of vigilance. However, only homogeneous groups experience an increase in fitness as group size increases. In contrast, vigilance behavior breaks down in larger, heterogeneous groups of semelparous prey. Error bars indicate bootstrapped 95\% confidence intervals over 100 replicates; some error bars are too small to be visible.}
\label{fig:grouped-treatment-comparison}
\end{center}
\end{figure*}

In our first experiment, all prey in the simulation are forced to forage in the same group, and the only trait that is evolving is the prey decision to be vigilant or not at every time step. Under these conditions, we find that prey living in homogeneous groups consistently evolve higher levels of vigilance than their counterparts living in heterogeneous groups (Figure~\ref{fig:grouped-treatment-comparison}). This suggests that organisms living in groups with high genetic relatedness are more likely to evolve cooperative strategies. Thus, in our model as in many natural systems, gregarious foraging is most favorable when genetic interests are aligned.

Figure~\ref{fig:grouped-treatment-comparison} also shows that semelparous prey are more likely to evolve vigilant strategies than iteroparous prey. This finding follows from the fact that semelparity selects more strongly for successful evasion of predator attacks, since prey death negates all previous foraging efforts. This effect is seen across both homogeneous and heterogeneous groups, indicating that semelparity is a strong enough selective pressure to act independently of group genetic composition. Importantly, prey vigilance does not evolve at all in the absence of predation (Figure S1), and gradually reducing the predation rate leads to a correspondingly gradual decrease in prey vigilance levels (Figure S2). Therefore, we know that the selection pressure imposed by predation is the primary driving force behind this evolved vigilance behavior.

All three treatments that evolve any level of vigilance also see the prevalence of vigilance decrease as group size increases. This pattern is important because it matches the pattern predicted by the many eyes hypothesis: As group size increases, individuals are able to rely more on collective rather than individual vigilance and can in turn devote more of their own time to foraging. Since we use a relative attack rate that scales the predator's attack frequency with group size, we can be sure that this phenomenon is due to group vigilance and not the dilution effect (fewer attacks per individual in larger groups) cited in other studies. We note that vigilance in the heterogeneous/semelparous treatment appears to evolve away almost entirely in a group size of 50. To explain why this trend might be due to something other than collective vigilance, we can instead look at trends in the fitness of the populations.

\begin{figure*}
\begin{center}
\includegraphics[width=0.95\textwidth]{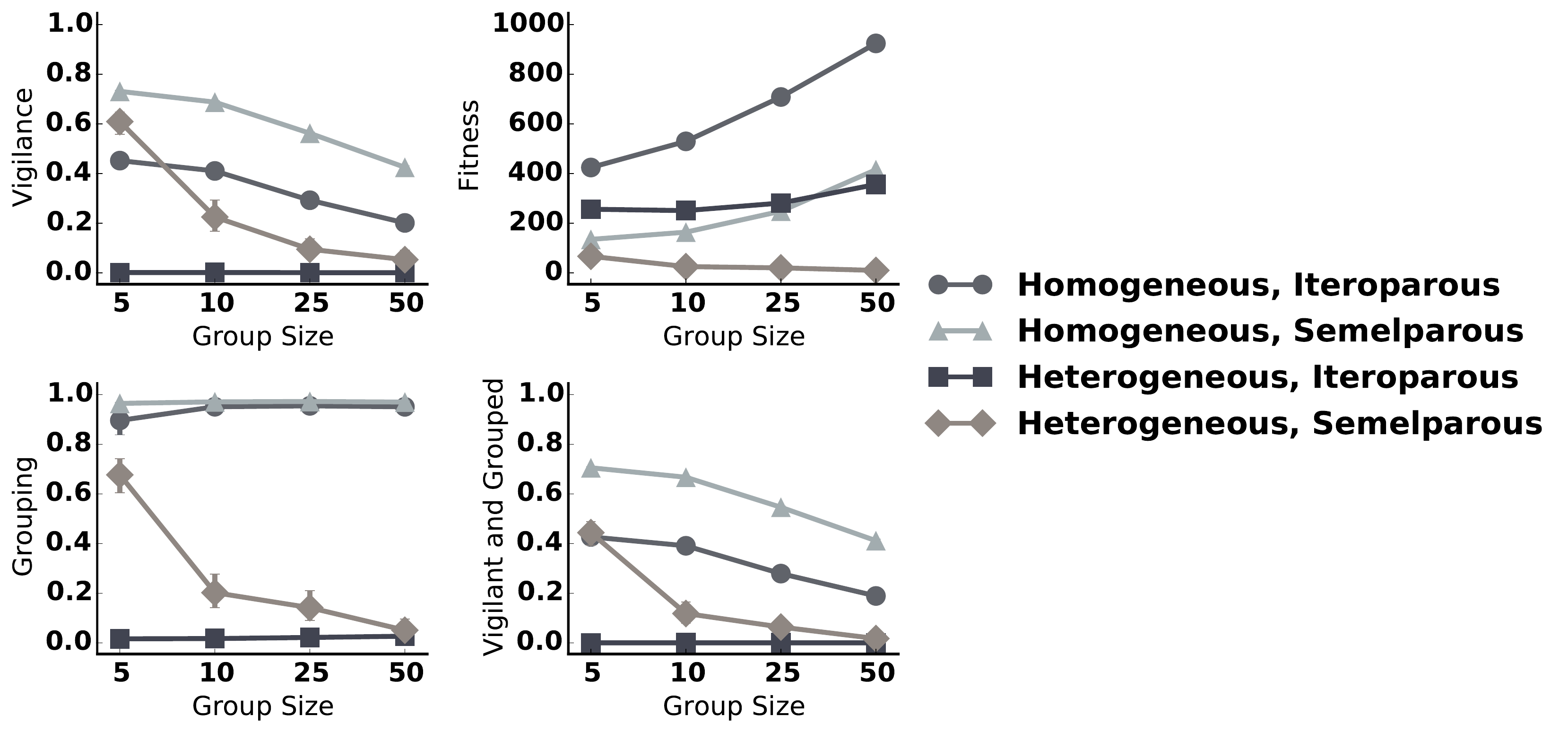}
\caption{{\bf Treatment comparison when prey can choose to forage in groups.} Allowing prey to decide whether they wish to be in the group produces very similar results compared to when they are forced to group. In homogeneous groups, prey choose to spend most of their time in the group. However, grouping breaks down (alongside vigilance) in heterogeneous groups of semelparous prey. This occurs despite there being no direct penalty assessed for choosing to group. Error bars indicate bootstrapped 95\% confidence intervals over 100 replicates; some error bars are too small to be visible.}
\label{fig:optional-treatment-comparison}
\end{center}
\end{figure*}

Several interesting trends are seen when we look at the influence of group size on average group fitness. In both homogeneous treatments, there is a steady increase in fitness with increasing group size, suggesting that gregarious foraging behavior is under positive selection. We see no significant fitness increase with group size in the heterogeneous/iteroparous populations, where the populations do not evolve vigilance behavior (Wilcoxon rank-sum $p=0.79$ between group size 5 and 50). Unlike the other treatments, the heterogeneous/semelparous populations actually experience a {\em decrease} in fitness with increasing group size (Wilcoxon rank-sum $p=2.77\times10^{-6}$ between group size 5 and 50), which suggests that cooperative behavior is not evolutionarily stable in larger heterogeneous groups. Accordingly, these findings highlight the fact that heterogeneous populations are much more susceptible to non-vigilant, ``cheating'' prey that sweep the population and reduce the overall population fitness.

\subsection*{Optional grouping}

So far we have shown that prey appear to take advantage of collective vigilance to increase their fitness when they are forced to group. We might expect from this (and the many eyes hypothesis predicts) that grouping provides a selective advantage. To test this expectation explicitly, we relax the constraints of the previous experiment by allowing the prey to evolve whether to group or not at every simulation time step. Since there is no direct fitness trade-off for grouping in this model yet (as there was for foraging and vigilance), this allows us to study whether the evolutionary advantages of grouping are favorable enough for vigilance and grouping to co-evolve.


Figure~\ref{fig:optional-treatment-comparison} shows that when we allow prey to choose to group, we find nearly the same results as before. This suggests that collective vigilance provides enough of a selective advantage to favor the evolution of grouping. It is not surprising that the homogeneous treatments evolve to group nearly 100\% of the time, given that the population is genetically identical and any ``altruistic'' action indirectly benefits the altruist as well. As in the forced grouping experiment, we observe a decline in fitness in the heterogeneous/semelparous populations as group size increases, to the point that the population is nearly driven extinct. The inability of the heterogeneous/semelparous populations to evolve consistently high levels of vigilance further supports the hypothesis that evolution is favoring short-term competitive advantages over long-term survival. This phenomenon is commonly known as the tragedy of the commons~\cite{Rankin2007,Wenseleers2004}, where selfish actions that provide an individual short-term benefit lead to a decrease in overall group fitness.

\subsection*{Tragedy of the commons in heterogeneous groups}

\begin{figure*}
\begin{center}
\includegraphics[width=0.95\textwidth]{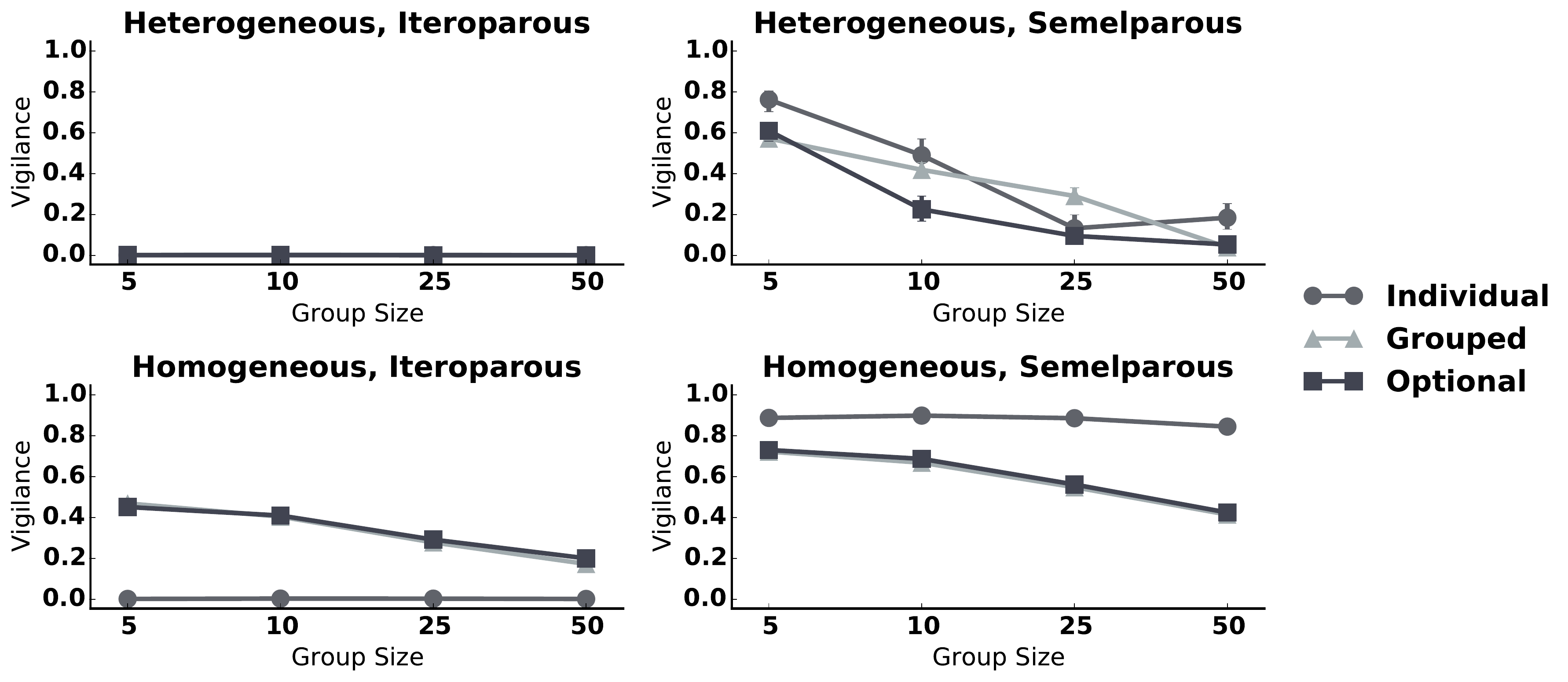}
\caption{{\bf Vigilance in prey with and without the option to forage in groups.} In homogeneous groups, prey with forced and optional grouping evolve similar vigilance behaviors. In contrast, individualistic (non-grouping) prey evolve vigilance behaviors that maximize individual fitness. Meanwhile, individuals in heterogeneous/semelparous populations with the option to group evolve to be less vigilant than either of the other two treatments. Error bars indicate bootstrapped 95\% confidence intervals over 100 replicates; some error bars are too small to be visible.}
\label{fig:tragedy-commons-vigilance}
\end{center}
\end{figure*}

To explore this apparent tragedy of the commons scenario further, we directly compare vigilance and fitness values from the forced and optional grouping experiments alongside a third experiment where we force the population to forage and survive as individuals. Figure~\ref{fig:tragedy-commons-vigilance} shows that when given the choice to group in the homogeneous treatments, prey behavior closely mirrors the behavior observed when forced to forage in a group. 
This observation confirms the previous suggestion that collective vigilance in homogeneous groups provides a fitness benefit that positively selects for gregarious foraging behaviors.

\begin{figure*}
\begin{center}
\includegraphics[width=0.6\textwidth]{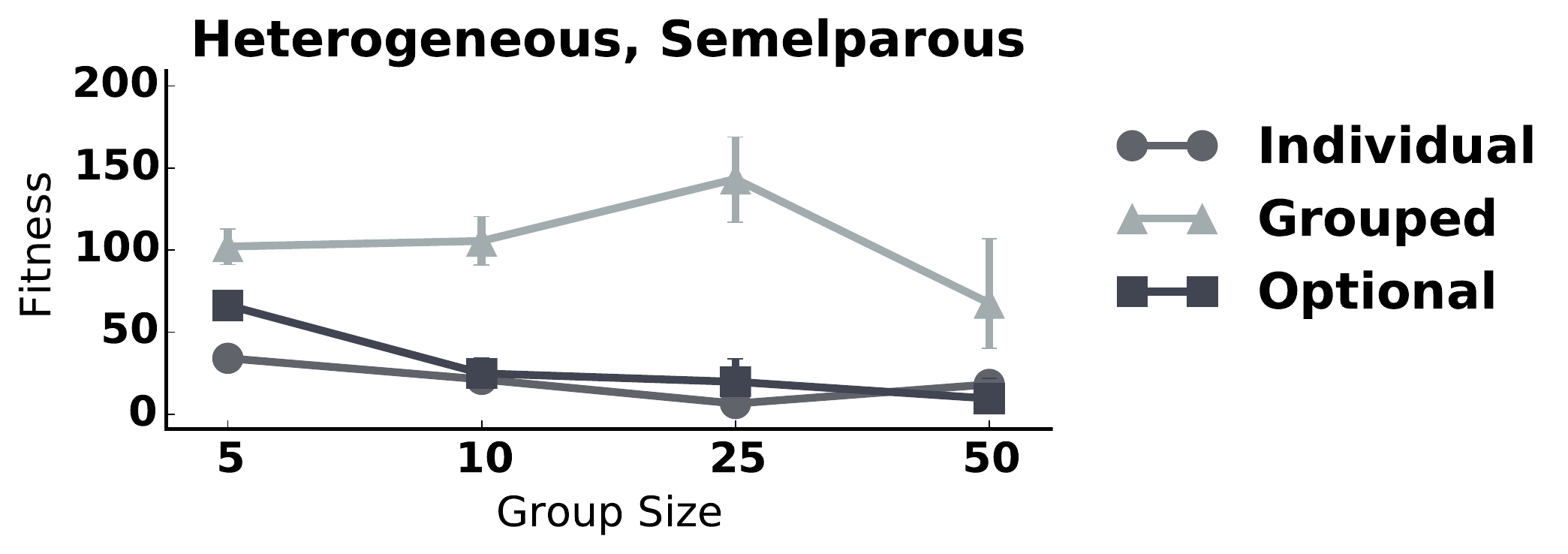}
\caption{{\bf Fitness for prey with and without the option to forage in groups.} In heterogeneous/semelparous groups, prey with the option to group have lower fitness than prey that are forced to group. Error bars indicate bootstrapped 95\% confidence intervals over 100 replicates; some error bars are too small to be visible.}
\label{fig:tragedy-commons-fitness}
\end{center}
\end{figure*}

In contrast to the homogeneous populations, heterogeneous populations are much less likely to evolve gregarious foraging behaviors. Heterogeneous/iteroparous populations never evolve vigilance behavior regardless of whether the prey are forced to group or not (Figure~\ref{fig:tragedy-commons-vigilance}). Similarly, heterogeneous/semelparous populations only evolve vigilance behavior in smaller groups, whereas the advantage of collective vigilance is lost in larger groups. At larger group sizes, prey with the ability to choose whether or not to forage in heterogeneous/semelparous groups instead evolve lower levels of vigilance than required to protect the group (Figure~\ref{fig:tragedy-commons-vigilance}), which results in a decrease in overall group fitness relative to prey that always forage in groups (Figure~\ref{fig:tragedy-commons-fitness}).

\subsection*{Explicit cost of grouping}

\begin{figure*}
\begin{center}
\includegraphics[width=0.9\textwidth]{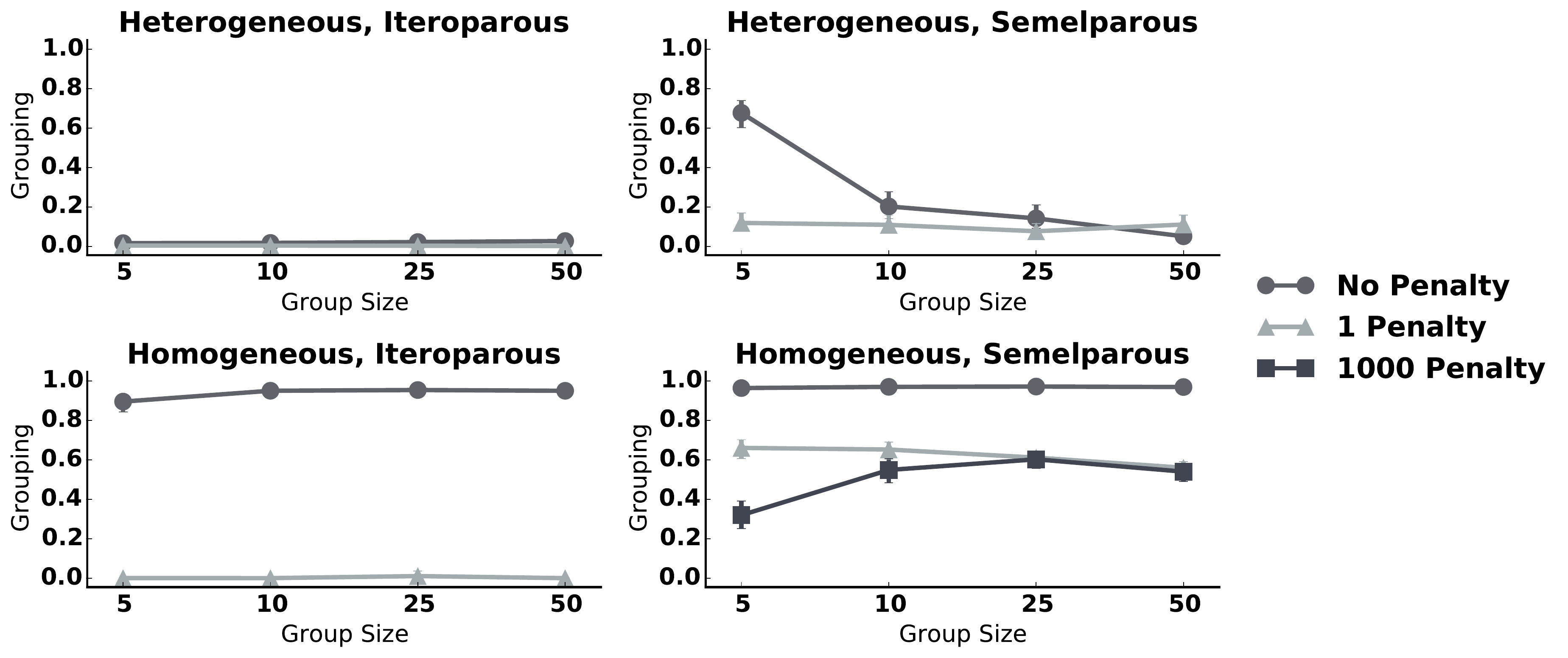}
\caption{{\bf Grouping behaviors in prey experiencing grouping penalties.} Even with a small grouping penalty ($M = 1.0$), all treatments except homogeneous/semelparous no longer evolve grouping behavior. Prey in the homogeneous/semelparous treatment only evolve slightly lower levels of grouping behavior, even with extreme penalties to foraging in a group ($M = 1,000$). Error bars indicate bootstrapped 95\% confidence intervals over 100 replicates; some error bars are too small to be visible.}
\label{fig:grouping-penalty-grouping}
\end{center}
\end{figure*}

In our final treatment, we investigate the impact of assessing a direct cost of foraging in a group (e.g., competition for food). Figure~\ref{fig:grouping-penalty-grouping} shows that except in the homogeneous/semelparous treatment, an explicit grouping cost selects against gregarious foraging behavior even when the grouping penalty is small ($M = 1.0$). Conversely, prey in the homogeneous/semelparous treatment maintain some level of gregarious foraging behavior even when the penalty for foraging in groups is extreme ($M = 1,000$). Therefore, we conclude that in the presence of even a small penalty for foraging in a group and the absence of additional selection pressures that favor gregarious foraging (e.g., improved social status for sentinels), only the combination of high genetic relatedness within the group and a semelparous reproductive strategy select strongly enough for gregarious foraging behavior to evolve.

\section{Discussion}

We found that gregarious foraging behavior can emerge under a variety of conditions when there is a benefit of vigilance and the spreading of information about predators. Prey that forage in homogeneous groups are more likely to evolve gregarious foraging behaviors compared to the those in heterogeneous groups. The same is true for semelparous organisms (who reproduce only once before death) compared to their iteroparous counterparts (who reproduce continually), but group homogeneity selects much more strongly for gregarious foraging behavior. 

Clearly, there are numerous challenges to evolving any form of cooperative behavior in a population with unconstrained genetic relatedness. However, we have shown here that when there is strong selection for survival (as in the heterogeneous/semelparous treatment), the benefit of information sharing via being vigilant and making alarm signals is sufficient to select for cooperative behavior in heterogeneous groups. This finding demonstrates that kinship is not necessary for cooperative behavior to evolve as long as there is some benefit to information sharing within the group, e.g., reducing predator attack efficiency.


Further, our results point to a heretofore unsuspected cost of gregarious foraging that is unique to heterogeneous groups. We call this the ``two-fold cost of vigilance.'' In our model, vigilance behavior in heterogeneous groups is more than a trade-off with foraging on the individual level. By choosing to be vigilant, prey also risk aiding in the survival of rival prey, which then puts the vigilant prey at a fitness disadvantage because it sacrificed a round of foraging to aid the rival prey. Together, these costs could explain why prey in heterogeneous groups evolve to be less vigilant than those in homogeneous groups.

At the same time, it is also possible that there are some evolutionary advantages unique to heterogenous groups that we have not yet addressed. For example, our model does not currently allow for any kind of specialization in roles between individuals, which could explain the presence of multi-species groups in nature~\cite{Goodale2005,Sridhar2009}. If the prey could evolve to preferentially pay attention to certain ``sentinel'' members of the population (who, in turn, choose to be vigilant nearly always in order to receive some form of rewards, e.g., food or increased social status) then perhaps an evolutionarily stable form of gregarious foraging could be found in heterogeneous groups of all sizes. It is even possible that such a complex social structure could out-perform the relatively primitive cooperation in our homogeneous groups.

Alongside genetic relatedness, another positive selective pressure for the evolution of vigilance is a semelparous reproductive strategy. When prey must survive any and all predator attacks in order to reproduce, the impetus to be vigilant is much greater. Semelparous organisms are known to be more risk-averse than similar, iteroparous organisms~\cite{Abrams1991}, and the decision to forage instead of being vigilant is an example of one such risky behavior. Thus, rather than spending most of their time foraging (as iteroparous prey evolve to do in our model), semelparous prey in our model tend to devote most of their time to watching for predators. When given the opportunity to group with other prey and take advantage of collective vigilance, semelparous prey are actually able to spend less time being vigilant. Thus, when semelparous prey evolve lower levels of vigilance in larger groups, we are observing the effect of collective vigilance.

Although our results suggest that the risk-averseness of semelparity induces semelparous prey to evolve to take advantage of collective vigilance, this selective pressure does not appear to be as strong as the pressure we observed in homogeneous groups. Proof of this observation can be found in the heterogeneous/semelparous treatment, where most group members attempt to cheat their way into collective vigilance by evolving lower levels of vigilance behavior than is observed in populations where prey are either forced to forage on their own or in the group (Figure~\ref{fig:tragedy-commons-vigilance}). Ultimately, this selfish behavior results in lower fitness than the fitness of prey that are forced to forage in groups (Figure~\ref{fig:tragedy-commons-fitness}), but the constantly-present, short-term benefits of selfishness appear to be too enticing to allow a more advantageous, cooperative behavior to emerge.

The breakdown of cooperation in the heterogeneous/semelparous populations suggests that the populations are succumbing to a tragedy of the commons~\cite{Rankin2007,Wenseleers2004}. In our experiments, all prey are competing against each other to forage as much food as possible without being captured by the predator. However, because there is an unlimited amount of food, the only depletable group resource is vigilance, which protects the entire group from the predator. As the resulting non-cooperative behavior in the heterogeneous/iteroparous populations demonstrate, absent any major selective pressures for collective vigilance, prey will evolve to selfishly forage 100\% of the time. Therefore, group homogeneity and semelparity correspond to two previously-established mechanisms for preventing a tragedy of the commons, namely kin selection and punishment for non-cooperative behaviors, respectively~\cite{Rankin2007}. The relative efficacy of these mechanisms to prevent cheating merits further investigation. For example, does group homogeneity play a larger role than reproductive strategy in the evolution of collective vigilance?

In the presence of even a small penalty for foraging in groups, we observe that only prey in homogeneous groups with a semelparous reproductive strategy are capable of evolving gregarious foraging behavior (Figure~\ref{fig:grouping-penalty-grouping}). This finding suggests that, in the absence of unlimited food resources or extreme predation rates, collective vigilance (i.e., the many eyes hypothesis) is insufficient to select for gregarious foraging. However, there may be important aspects of natural systems that select for gregarious foraging that we did not model here. For example, predators have been observed to preferentially attack non-vigilant prey in groups~\cite{Krause1996}, which would require prey to be vigilant even without the benefit of collective vigilance. Thus, it would be informative in future work to model such a preference for non-vigilant prey and observe the evolution of gregarious foraging under those conditions.

The experimental platform presented here enables a plethora of hypotheses to be studied in future work. Given that there is considerable evidence suggesting that foraging and vigilance behaviors are not mutually exclusive in some species~\cite{Lima1999,Cresswell2003,FernandezJuricic2008}, it would be instructive to relax that assumption in this model and make foraging only a ``reduced vigilance'' state. Furthermore, we assume here that prey cannot detect the size of their group; a useful extension would be to allow prey to detect their group's size and study collective vigilance in prey that evolved in varying group sizes. We also assume here that prey always communicate the presence of the predator to their group members. Given that it may not always be evolutionarily advantageous to aid other group members, another informative extension would be to allow prey to optionally make their alarm signals upon detection of the predator. Finally, there are several hypotheses other than the many eyes hypothesis that could be explored with a model similar to the one presented here, such as the predator confusion hypothesis~\cite{Olson2013PredatorConfusion} and the selfish herd hypothesis~\cite{Olson2013SelfishHerd,Olson2014SelfishHerd}. Once all of these hypotheses have been studied in isolation, we can then combine them into a single model to study their relative importance and separate the adaptive benefits (those that select for the evolution of grouping) from the chance side effects of grouping. Such experiments will be invaluable for understanding how and why animals evolve grouping behavior.

\section*{Acknowledgements}

We thank Art Covert and the Freshman Research Initiative at The University of Texas at Austin for their support through the development of this project. This research has been supported by the National Science Foundation BEACON Center under Cooperative Agreement DBI-0939454, and Michigan State University through computational resources provided by the Institute for Cyber-Enabled Research.

\bibliography{references}
\bibliographystyle{unsrt}

\end{document}